\documentclass[reprint,amsmath,amssymb,aps,pre,superscriptaddress,floatfix]{revtex4-2}

\usepackage{graphicx}
\usepackage{bm}
\usepackage[colorlinks=true,allcolors=blue]{hyperref}

\newcommand{\Lf}{L_{\mathrm{f}}}
\newcommand{\Psic}{\psi_{\mathrm{c}}}
\newcommand{\kT}{k_{\mathrm{B}}T}
\newcommand{\Uc}{U_{\mathrm{c}}}
\newcommand{\dd}{\mathrm{d}}

\begin{document}

\title{A statistical-physics framework for translocation elastometry of deformable particles}

\author{Pierre Ronceray} \email{pierre.ronceray@univ-amu.fr}
\affiliation{Aix-Marseille Universit\'e, CNRS, CINaM, Turing Centre for Living Systems, Marseille, France}

\begin{abstract} A soft particle driven through a pore narrower than itself must deform to pass, and how quickly it does so is set by how hard it is to squeeze. We propose a mathematical framework for turning rate measurements of this driven, stochastic passage into quantitative mechanical measurements. Treating the entry of the particle as one-dimensional Brownian dynamics across an elastic barrier, we solve the transport problem exactly and identify two dynamical regimes: at low drive the passage is thermally activated and limited by the energy needed to deform the particle, and at high drive it is friction-limited. We propose a framework to extract the particle's deformation energy and relevant geometrical information by combining measurements in these two regimes. This method could be used in the context of nanopore sensing, where the drive is an applied voltage: the framework then provides a self-calibrating route---translocation elastometry---from a current--voltage measurement to the elasticity and shape of individual soft nanoparticles.
\end{abstract}

\maketitle

\section{Introduction}
The mechanical properties of micro- and nanoscale objects---the stiffness of a virus capsid, the
bending rigidity of a lipid vesicle, the elasticity of a cell---are increasingly recognised as
informative, label-free phenotypes~\cite{roos2010,otto2015}. Measuring them on individual, freely
suspended particles remains difficult: atomic-force nanoindentation resolves single capsids but
requires immobilisation and is low-throughput~\cite{roos2010}, while deformability cytometry
phenotypes cells at high rate but operates in the micron range~\cite{otto2015}. Solid-state
nanopores~\cite{dekker2007,wanunu2012,garaj2010,emmerich2024} occupy an attractive middle ground: a particle
driven electrophoretically through a pore narrower than itself must squeeze through, and the
kinetics of this forced passage are sensitive to how hard the particle is to compress. Nanopore and
constriction experiments have indeed observed the deformation of microgels, liposomes, and soft
nanoparticles during translocation~\cite{holden2011,darvish2015,chen2017,zhu2020}, and have begun to
infer particle stiffness by comparing the current-blockade signal to that of a rigid
reference~\cite{darvish2019}. These readouts rest on the amplitude and shape of the current blockade
rather than on the passage kinetics that carry the deformation energy.

Turning such observations into a \emph{quantitative} mechanical measurement requires a theory that
maps the accessible observable---the mean translocation velocity as a function of applied
voltage~\cite{storm2005,plesa2014}---onto the microscopic energy of deformation. Polymer
translocation offers a mature statistical-physics template, in which the passage is described as
Fokker-Planck motion of a reaction coordinate across a free-energy
barrier~\cite{sungpark1996,muthukumar1999,muthukumar2001,chuang2001,kantor2004,sakaue2007}; there, however,
the controlling free energy is \emph{entropic}. For a deformable particle the controlling energy is
instead \emph{elastic}, and the geometry of forced radial entry is qualitatively different, so the
theory must be built anew. Continuum simulations of elastic capsules and vesicles transiting
microfluidic constrictions resolve this deformation in detail~\cite{park2013,kahali2024}, but do so
deterministically at fixed drive; here we instead reduce the passage to a single stochastic
coordinate, which is what makes the driven \emph{kinetics} --- and through them the deformation
energy --- directly accessible.

We model the driven passage as one-dimensional overdamped motion
of the particle centre across a piecewise-linear elastic landscape---the \emph{funnel model}---set
by a deformation energy $\Delta$ and a funnel length $\Lf$, and solve the resulting steady-state
Fokker--Planck problem in closed form [Eq.~\eqref{eq:vfinal}]. The solution exhibits two
experimentally salient regimes, a low-drive \emph{activated} regime limited by the deformation
barrier and a high-drive \emph{frictional} regime, separated by a crossover force
$\Psic=\Delta/\Lf$ at which the entrance barrier vanishes. This structure carries a dual mechanical
readout: combining the two regimes returns the deformation energy $\Delta/\kT$ directly, eliminating
the unknown force-to-voltage conversion---a self-calibrating ``translocation elastometry''---while the
funnel length $\Lf$ encodes information about the particle's shape. We treat the system
generically---a deformable particle of radius $R$ driven through a pore of radius $a<R$---and
validate every result against direct Langevin simulation.

\section{The funnel model}\label{sec:model}
\begin{figure*}[tb]
  \centering
  \includegraphics[width=\textwidth]{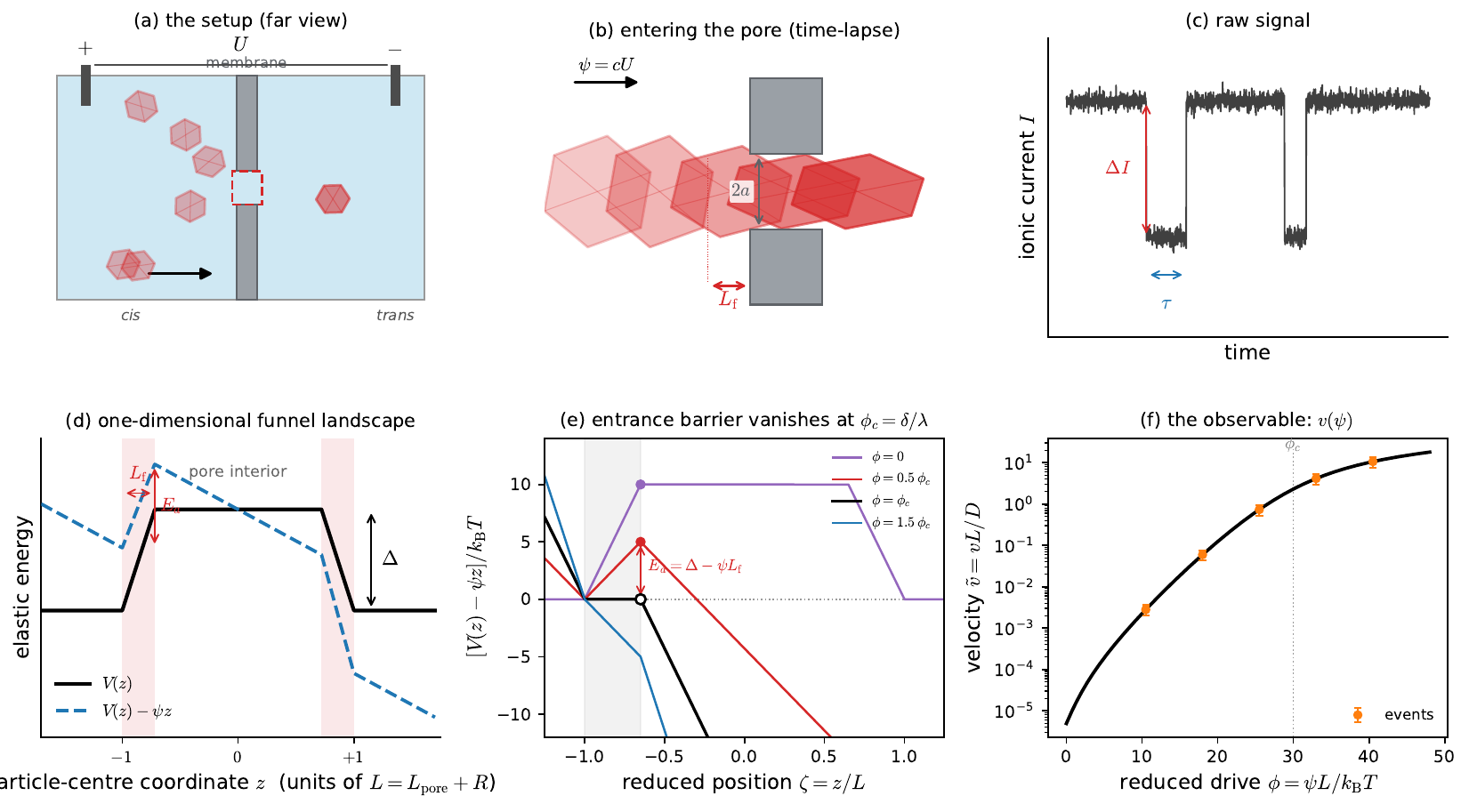}
  \caption{From experiment to model. (a) An example setting: soft particles suspended in an
  electrolyte are driven by a voltage $U$ across a membrane pierced by a single pore. (b) Time-lapse
  of a single particle entering the pore: driven by a force $\psi=cU$, a particle of radius $R>a$
  compresses radially to pass, over an axial funnel length $\Lf$---the distance its centre advances
  between first contact and full insertion. (c) The raw observable: each translocation is a blockade
  in the ionic current, of depth $\Delta I$ and dwell time $\tau$. (d) The one-dimensional reduction:
  the particle-centre coordinate $z$ experiences a piecewise-linear elastic potential $V(z)$ of depth
  $\Delta$, ramped over $\Lf$ at each mouth and tilted by the drive to $V(z)-\psi z$; the entrance
  barrier is $E_{\mathrm a}=\Delta-\psi\Lf$. (e) Tilting the landscape with increasing drive: the
  entrance barrier shrinks and vanishes at $\psi=\Psic=\Delta/\Lf$, where the entry funnel is flat and
  passage becomes barrierless (colours mark the dynamical regimes). (f) The mean translocation
  velocity $v\sim L/\tau$, collected over many events at several drives, is the observable this paper
  models (solid curve, exact closed form; points, representative events; $\phi_c$ marks the
  crossover).}
  \label{fig:schematic}
\end{figure*}

We describe the particle by the position $z$ of its centre along the pore axis. Deformation is
encoded in a $z$-dependent elastic energy $V(z)$, which renders the dynamics Markovian in this
single coordinate. The particle is undeformed (energy $0$) when fully outside the pore and maximally
deformed (energy $\Delta$) when fully inside; the energy rises over an axial distance $\Lf$, the
\emph{funnel length}, which measures how far the particle advances as it squeezes in. We take the
simplest landscape consistent with this picture, piecewise linear in $z$ (Fig.~\ref{fig:schematic}),
\begin{equation}
  V(z)=
  \begin{cases}
    0 & |z|>L,\\[2pt]
    \Delta\,(L-|z|)/\Lf & L-\Lf<|z|<L,\\[2pt]
    \Delta & |z|<L-\Lf,
  \end{cases}
  \label{eq:V}
\end{equation}
where $L=L_{\mathrm{pore}}+R$ is the half-length over which the particle interacts with the pore.
The particle undergoes overdamped Langevin dynamics~\cite{gardiner2009} with diffusion constant $D$ and mobility
$\mu=D/\kT$ (fluctuation--dissipation), driven by a constant force $\psi$, the \emph{drive}. We keep
the drive abstract; in the context of applications to nanopores it would originate from an applied voltage (Fig.~\ref{fig:schematic}a), and we return
to that coupling below.
The force is
\begin{equation}
  F(z)=-V'(z)+\psi=
  \begin{cases}
    \psi-\Psic & \text{funnel (in)},\\
    \psi & \text{inside},\\
    \psi+\Psic & \text{funnel (out)},
  \end{cases}
  \label{eq:force}
\end{equation}
with the \emph{crossover force}
\begin{equation}
  \Psic\equiv\Delta/\Lf .
  \label{eq:psic}
\end{equation}
The decomposition~\eqref{eq:force} makes the meaning of $\Psic$ transparent. In the entrance funnel
the elastic potential rises with axial slope $\Delta/\Lf=\Psic$, so the net force there is
$\psi-\Psic$: the drive must overcome the funnel slope, not merely the average slope $\Delta/L$
across the whole interaction region. The name reflects this picture: the pore mouth acts as a
funnel that progressively compresses the particle over the length $\Lf$, and it is the \emph{slope}
of that funnel, $\Delta/\Lf$, rather than the total energy $\Delta$, that sets the force scale for
entry. A longer funnel is a gentler ramp---at fixed deformation energy it lowers both the crossover
force $\Psic=\Delta/\Lf$ and the entrance barrier below---so a given drive carries the particle
through more easily. The height of the
entrance barrier a particle must cross is
\begin{equation}
  E_{\mathrm a}(\psi)=\Delta-\psi\Lf ,
  \label{eq:barrier}
\end{equation}
which decreases with drive and vanishes \emph{exactly} at $\psi=\Psic=\Delta/\Lf$: above this force
the entrance funnel tilts downhill and there is no barrier to activate over. Thus $\Psic$ is not a
dimensional estimate but the genuine drive at which activated entry gives way to barrierless,
frictional passage. The fact that a stronger drive should lower and eventually overcome the entry barrier has
been noted qualitatively in translocation experiments~\cite{zhu2020}. Because $\Psic$ carries the
funnel length, the crossover force is itself geometry-dependent, which is implies that some properties of the  particle's
shape are observable in the dynamics (Sec.~\ref{sec:geometry}).

In the nanopore realisation the drive is set by the electrode voltage $U$, but the relationship
between $\psi$ and $U$ is not simple: electrostatic forces act mostly to capture particles at the
mouth, leaving an unknown residual force during passage that need not equal the bare
electrophoretic force~\cite{keyser2006,qiao2019}; continuum models of
deformable-particle translocation resolve this force numerically~\cite{he2021}. We therefore write
$\psi=cU$ with an unknown force-to-voltage conversion $c$ and return to its elimination in
Sec.~\ref{sec:elastometry}. The framework itself needs only that the drive be constant and tunable;
other realisations (mechanical or osmotic driving through a constriction) map on identically.

Particles are supplied on one side and absorbed on the other, giving boundary densities
$p(-L)=\rho$ and $p(+L)=0$. In steady state the probability density $p(z)$ carries a uniform
flux $J$ obeying the Fokker--Planck relation $\mu\,p(z)\,F(z)=J+D\,p'(z)$~\cite{risken1989}. The observable of
interest is the mean translocation velocity
\begin{equation}
  v\equiv J/\rho ,
  \label{eq:vdef}
\end{equation}
which eliminates the concentration- and drive-dependent prefactor $\rho$.

\section{Exact solution}\label{sec:exact}
\begin{figure*}[tbp]
  \centering
  \includegraphics[width=\textwidth]{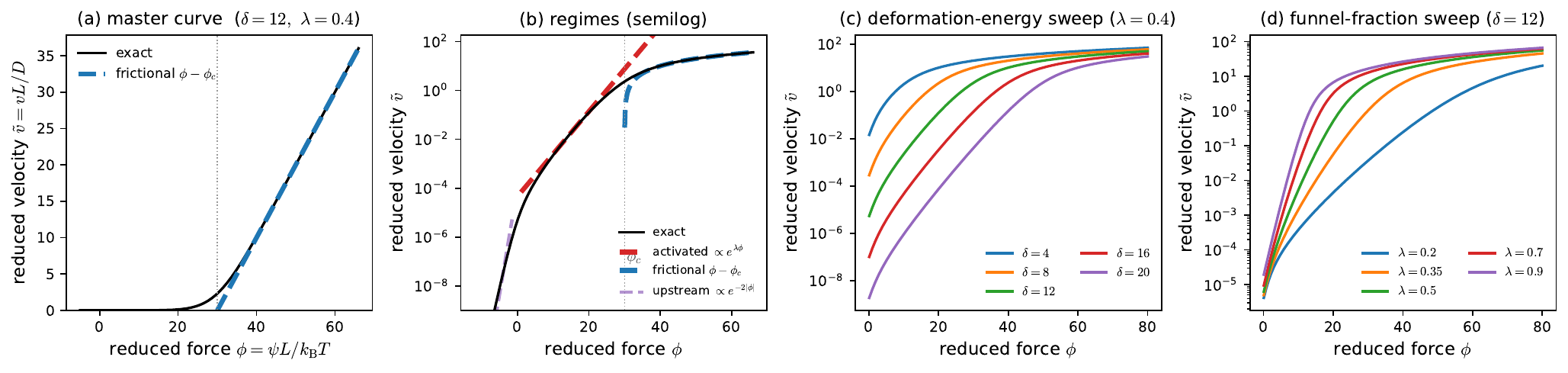}
  \caption{Master velocity curve. (a) Reduced velocity $\tilde v=vL/D$ versus reduced force
  $\phi=\psi L/\kT$ on a linear scale, with the frictional asymptote $\phi-\phi_c$. (b) The same on a
  semilog scale, with the activated and upstream asymptotes and the crossover $\phi_c$ marked.
  (c,d) Sweeps over the reduced deformation energy $\delta=\Delta/\kT$ and funnel fraction
  $\lambda=\Lf/L$. All curves are the exact closed form Eq.~\eqref{eq:vfinal}.}
  \label{fig:master}
\end{figure*}

Equation~\eqref{eq:vdef} with linear-by-parts force defined in Eq.~\eqref{eq:force} can be solved in closed form. Introducing the integrating factor
$I(z)=\exp\!\big([V(z)-(L+z)\psi]/\kT\big)$, the steady solution with $p(+L)=0$ is
$p(z)=-\frac{J}{D\,I(z)}\int_L^z I(s)\,\dd s$, and evaluating at $z=-L$ gives $\rho$. Performing the
three region integrals yields
\begin{widetext}
\begin{equation}
  v=\mu\left[
  \underbrace{\frac{1}{\psi-\Psic}}_{\text{frictional}}
  +\frac{\Psic}{\psi}\,e^{(\Psic-\psi)\Lf/\kT}
  \left(\underbrace{\frac{1}{\Psic-\psi}}_{\text{activated}}
  -\frac{e^{-2\psi(L-\Lf)/\kT}}{\Psic+\psi}\right)
  -\underbrace{\frac{e^{-2\psi L/\kT}}{\Psic+\psi}}_{\text{upstream}}
  \right]^{-1}.
  \label{eq:vfinal}
\end{equation}
\end{widetext}
The three labelled groups anticipate the dynamical regimes below. It is convenient to nondimensionalise,
measuring energies in $\kT$ and lengths in $L$:
\begin{equation}
  \delta=\frac{\Delta}{\kT},\quad \lambda=\frac{\Lf}{L},\quad \phi=\frac{\psi L}{\kT},\quad
  \phi_c=\frac{\delta}{\lambda}=\frac{\Psic L}{\kT}.
  \label{eq:nondim}
\end{equation}
The reduced velocity $\tilde v\equiv vL/D=1/B(\phi;\delta,\lambda)$ is the inverse of a dimensionless
bracket $B$ assembled from the same three region integrals; it is finite everywhere, the apparent
poles of Eq.~\eqref{eq:vfinal} at $\phi=\pm\phi_c,0$ being removable. We verify symbolically that the
bracket equals the first-principles integral of $I(z)$ and that Eq.~\eqref{eq:vfinal} reproduces it
(Sec.~\ref{sec:numerics}). Representative curves are shown in Fig.~\ref{fig:master}.

\section{Dynamical regimes}\label{sec:regimes}
\begin{figure}[tbp]
  \centering
  \includegraphics[width=\columnwidth]{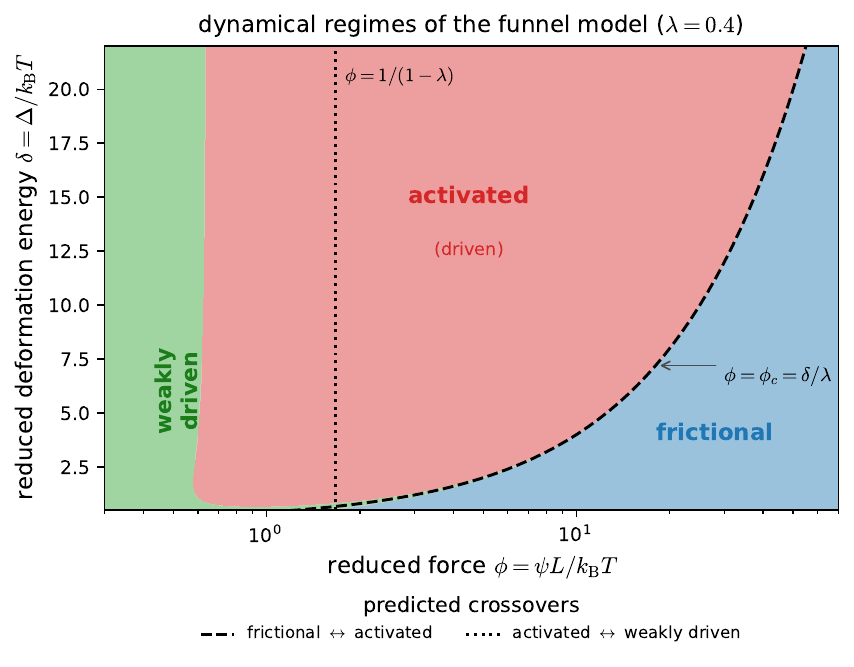}
  \caption{Dynamical-regime phase diagram in the reduced force--deformation-energy plane
  $(\phi,\delta)$ at fixed $\lambda$. Colours mark which asymptote best approximates the exact
  velocity; the dashed/dotted lines are the predicted crossovers $\phi=\phi_c=\delta/\lambda$ and
  $\phi=1/(1-\lambda)$.}
  \label{fig:phase}
\end{figure}

We focus on the strongly deforming case $\delta\gg1$. Expanding Eq.~\eqref{eq:vfinal} identifies the
regimes (Fig.~\ref{fig:master}); as the drive grows the entrance barrier $E_{\mathrm a}=\Delta-\psi\Lf$
shrinks and is cancelled at $\psi=\Psic$ (Fig.~\ref{fig:schematic}e), and the dominant balance shifts. A
phase diagram in the $(\phi,\delta)$ plane is shown in Fig.~\ref{fig:phase}.

\paragraph{Frictional regime ($\psi\Lf\gg\Delta$).} The first term dominates and
\begin{equation}
  v\simeq\mu\,(\psi-\Psic),
  \label{eq:friction}
\end{equation}
a linear force--velocity law whose slope gives the mobility and whose intercept gives the crossover
force $\Psic=\Delta/\Lf$.

\paragraph{Activated regime ($\psi\Lf\ll\Delta$, $\psi(L-\Lf)\gg\kT$).} Entry is rate-limited by the
deformation barrier while transport inside is ballistic---a driven Kramers escape over the elastic
barrier~\cite{kramers1940,hanggi1990}---giving
\begin{equation}
  v\simeq\mu\,\psi\Big(1-\frac{\psi}{\Psic}\Big)\,e^{-(\Psic-\psi)\Lf/\kT},
  \label{eq:activated}
\end{equation}
so that on a semilog plot
\begin{equation}
  \ln v\simeq \frac{\psi\Lf}{\kT}+\text{const},
  \label{eq:logslope}
\end{equation}
with a slope that reads off $\Lf/\kT$.

\paragraph{Other regimes.} For weak drive ($\psi L\ll\kT$) the velocity saturates at a
concentration-driven, drift-independent value $v\sim\mu\,e^{-\Delta/\kT}/[2(\Psic^{-1}+(L-\Lf)/\kT)]$;
for strong reverse drive ($-\psi L\gg\kT$) a residual upstream current survives,
$v\sim e^{-2|\psi|L/\kT}$. These regimes are induced by the choice of boundary conditions and, for $\Delta \gg 1$, would likely not be observable.

Each asymptote follows from Eq.~\eqref{eq:vfinal} in the corresponding limit, and together they tile
the phase diagram of
Fig.~\ref{fig:phase}, whose boundaries $\phi=\phi_c$ (frictional/activated) and $\phi\sim 1/(1-\lambda)$
(activated/weak) we obtain analytically and confirm by classifying the exact solution.

\section{Geometric interpretation of the funnel length}\label{sec:geometry}
\begin{figure*}[tbp]
  \centering
  \includegraphics[width=\textwidth]{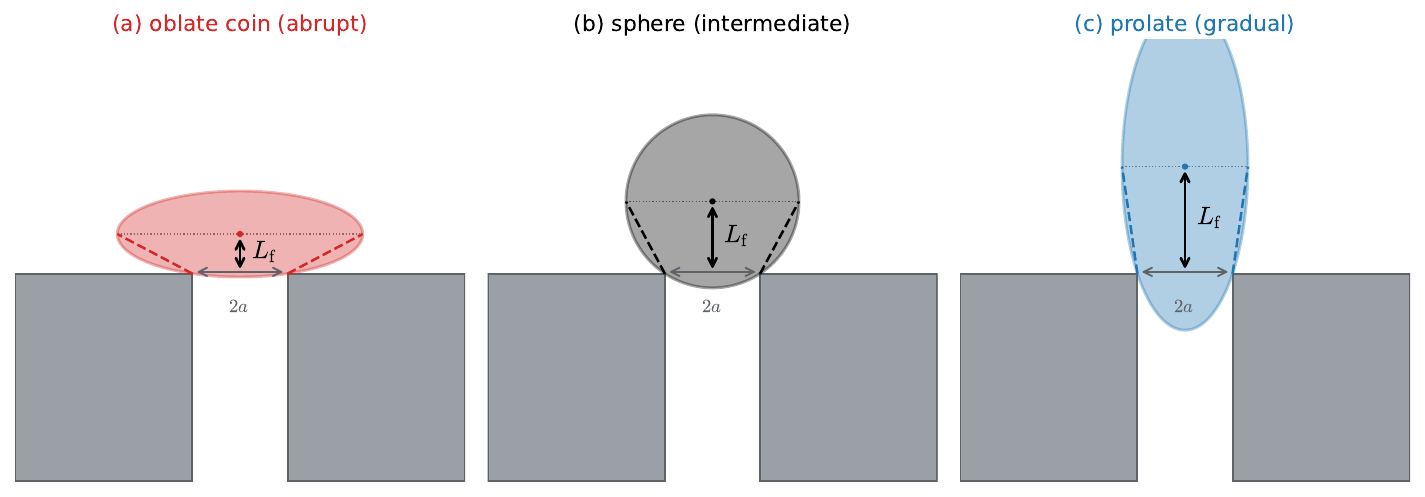}
  \caption{The funnel length carries geometric information. Three equal-volume particles of
  different shape driven along the pore axis into the same pore, drawn at first contact so the funnel
  length is directly comparable: (a) an oblate ``coin'' funnels in abruptly (small $\Lf$), (b) a
  sphere intermediately, and (c) a prolate body gradually (large $\Lf$). Dashed lines mark the
  effective funnel---the taper from pore rim to equator over the axial distance $\Lf$. For a
  spheroid of equatorial radius $b$ and polar semi-axis $c$, $\Lf=c\sqrt{1-(a/b)^2}$ (sphere:
  $\Lf=\sqrt{R^2-a^2}$), so $\Lf$ classifies the shape of the particle--pore pair and vanishes in the
  flat-blocker limit $c\to0$.}
  \label{fig:geometry}
\end{figure*}

The funnel length is a geometric property of the particle--pore pair. Model the particle as an
axisymmetric spheroid with equatorial radius $b$ and polar semi-axis $c$ along the pore axis (a sphere
has $b=c=R$), driven into a pore of radius $a<b$. When the centre sits a distance $s$ behind the pore
mouth, the particle's cross-section at the mouth has undeformed radius $b\sqrt{1-(s/c)^2}$. Compression
begins when this equals the pore radius, at $s_0=-c\sqrt{1-(a/b)^2}$, and is complete when the equator
reaches the mouth, at $s=0$. The axial distance over which the elastic energy is built up is therefore
\begin{equation}
  \Lf=c\sqrt{1-(a/b)^2}\qquad\xrightarrow{\text{sphere}}\qquad \Lf=\sqrt{R^2-a^2}.
  \label{eq:Lf-geom}
\end{equation}
This has the expected limits: $\Lf\to0$ as $a\to b$ (an abrupt entry
for a barely-confining pore) and $\Lf\to c$ as $a\to0$. Shape enters
through the ratio $c/b$: at fixed volume, an oblate ``coin'' entering
face-on ($c<b$) funnels in over a short axial distance, a sphere
intermediately, and a prolate body gradually
(Fig.~\ref{fig:geometry}). In the extreme oblate limit the particle
becomes a flat blocker that seals the pore without a smooth entrance,
$\Lf\to0$: the entry is a step rather than a funnel and the activation
energy becomes drive-independent---the frictional regime
disappears. The funnel length is thus a shape classifier of the
particle--pore pair, and reading it off the dynamics is complementary
to the deformation energy: where $\Delta$ fixes the mechanical scale
(Sec.~\ref{sec:modulus}), $\Lf$ reports on geometry. This is the dual
readout that makes the measurement informative---the same
velocity--voltage curve constrains both how soft the particle is,
through $\Delta$, and what shape it is, through $\Lf$.

Two effects enrich this geometric picture. First, for a nearly incompressible particle the radial
squeeze forces an axial elongation, lengthening $\Lf$ by a volume-conservation factor that grows with
the Poisson ratio. Second, and more consequentially for the measurement, the funnel length depends on
the \emph{orientation} in which the particle meets the pore. A non-spherical particle arriving along
its short axis presents a different leading cross-section---and therefore a different $\Lf$, through the
effective $c$ in Eq.~\eqref{eq:Lf-geom}---than the same particle arriving along its long axis. A
population of identical particles thus samples a distribution of arrival orientations, and the measured
signal is an average of the single-passage dynamics over this distribution,
\begin{equation}
  \langle v(\psi)\rangle=\int v\!\left(\psi;\Delta(\Omega),\Lf(\Omega)\right)\,p(\Omega)\,\dd\Omega,
  \label{eq:angle-average}
\end{equation}
with $p(\Omega)$ the arrival-orientation distribution set by the incoming flux and any
pre-alignment in the approaching field; orientation-dependent passage of this kind has been observed
for non-spherical objects, whose translocation signatures sort into distinct orientational
classes~\cite{zhu2020}. Except for a sphere, the observed curve is therefore not that
of a single sharp funnel but an orientation-averaged one, whose effective entrance is smoother than any
individual passage. This orientation average is one physical origin of the smooth-funnel
\emph{misspecification} we analyse in Sec.~\ref{sec:validation}: fitting the sharp piecewise-linear
model to an intrinsically averaged signal biases the recovered parameters, and motivates the
smooth-contact refinement developed there.

\section{From deformation energy to elastic modulus}\label{sec:modulus}
The deformation energy sets the absolute mechanical scale. For a linear-elastic solid the energy to
compress the particle to the pore radius scales, by dimensional analysis, as
\begin{equation}
  \Delta = Y\,R^3\, g(a/R;\nu),
  \label{eq:delta-Y}
\end{equation}
with $Y$ the Young's modulus, $\nu$ the Poisson ratio, and $g$ a dimensionless function of the
geometric ratio. In the small-compression (Hertzian contact) limit $g\propto (1-a/R)^{5/2}/(1-\nu^2)$,
i.e. $\Delta\sim [Y/(1-\nu^2)]\,R^{1/2}(R-a)^{5/2}$~\cite{hertz1882,landau-elasticity}; at larger
strain a harmonic estimate gives $\Delta\sim Y R\,(R-a)^2$. For a thin shell or capsule the relevant
modulus is the bending rigidity $\kappa$, and $\Delta\simeq\kappa\,h(a/R)$ with $h\sim(R/a-1)^2$.
Inverting Eq.~\eqref{eq:delta-Y} with the measured $\Delta/\kT$ and known geometry yields the modulus,
$Y=(\Delta/\kT)\,\kT/[R^3 g(a/R;\nu)]$. The method is most informative in the
soft-particle regime, where the deformation energy is comparable to a few $\kT$: for a particle of
radius $R\simeq15$~nm entering a pore of radius $a\simeq10$~nm, a resolvable $\Delta/\kT\sim5$--$12$
corresponds to a Young's modulus of order $10^{5}$~Pa ($\sim0.1$~MPa), i.e. the range spanned by
microgels, emulsion droplets, and other soft colloids. Stiffer objects raise
$\Delta$ steeply---the deformation energy quickly exceeds the window over which translocation is both
thermally accessible and deformation-limited---so the readout is best suited to genuinely soft
particles rather than rigid capsids.

\section{Translocation elastometry}\label{sec:elastometry}
We now specialise the dual readout to the voltage-driven realisation, where the drive is set by the
applied voltage through $\psi=cU$; the same construction applies to any tunable drive.
Fitting the full $v(\psi)$ of Eq.~\eqref{eq:vfinal} to data is hampered by the unknown force-to-voltage
conversion $c$, by velocities spanning orders of magnitude, and by sparse, noisy sampling of the
informative low-drive (activated) regime. We circumvent these difficulties by combining the two
scaling laws. In the frictional regime a linear fit $v\simeq A\,(U-\Uc)$ gives the intercept
\begin{equation}
  \Uc=\Delta/(c\Lf),
  \label{eq:Uc}
\end{equation}
while in the activated regime a semilog fit $\ln v\simeq B\,U+\text{const}$ gives the slope
\begin{equation}
  B=c\Lf/\kT.
  \label{eq:B}
\end{equation}
Both depend on the nuisances $c$ and $\Lf$, but their product does not:
\begin{equation}
  \boxed{\,B\,\Uc=\Delta/\kT\,.}
  \label{eq:elastometry}
\end{equation}
The combination returns the dimensionless deformation energy directly, with no knowledge of the
force-to-voltage conversion or the funnel length---a self-calibrating readout. We note a useful
refinement: the activated semilog slope $\dd\ln v/\dd\phi=\lambda+1/\phi-1/(\phi_c-\phi)$ equals
exactly $\lambda$ (i.e. reads $\Lf/L$ without bias) at $\psi=\Psic/2$; fitting the slope about that
point minimises systematic error.

\section{Validation and parameter recovery}\label{sec:validation}
\begin{figure}[tbp]
  \centering
  \includegraphics[width=\columnwidth]{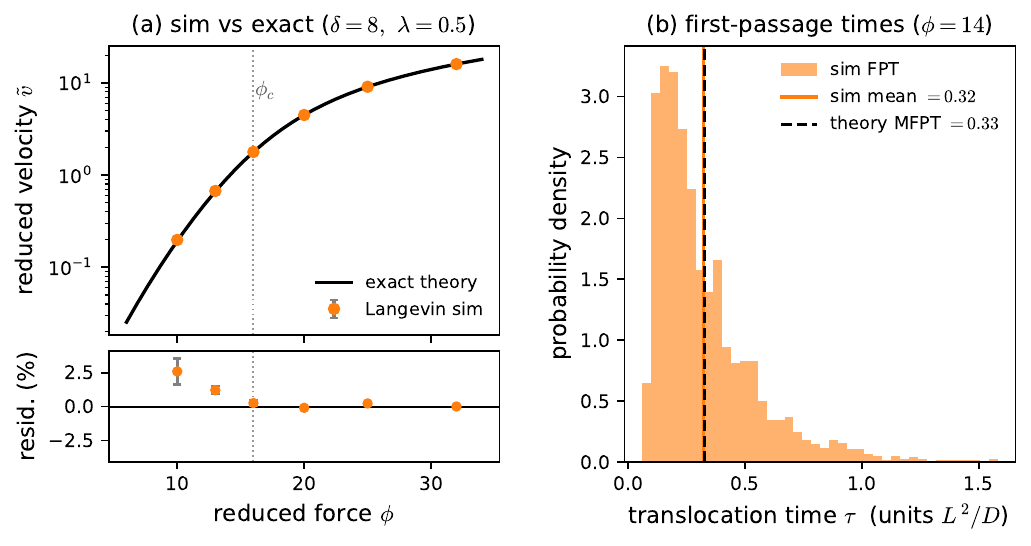}
  \caption{Langevin simulation validates the theory. (a) Source--sink simulated velocity (points)
  on the exact curve (line) across regimes. (b) Simulated translocation-time distribution with the
  simulated and analytic mean first-passage times.}
  \label{fig:sim}
\end{figure}

\begin{figure*}[tbp]
  \centering
  \includegraphics[width=0.8\textwidth]{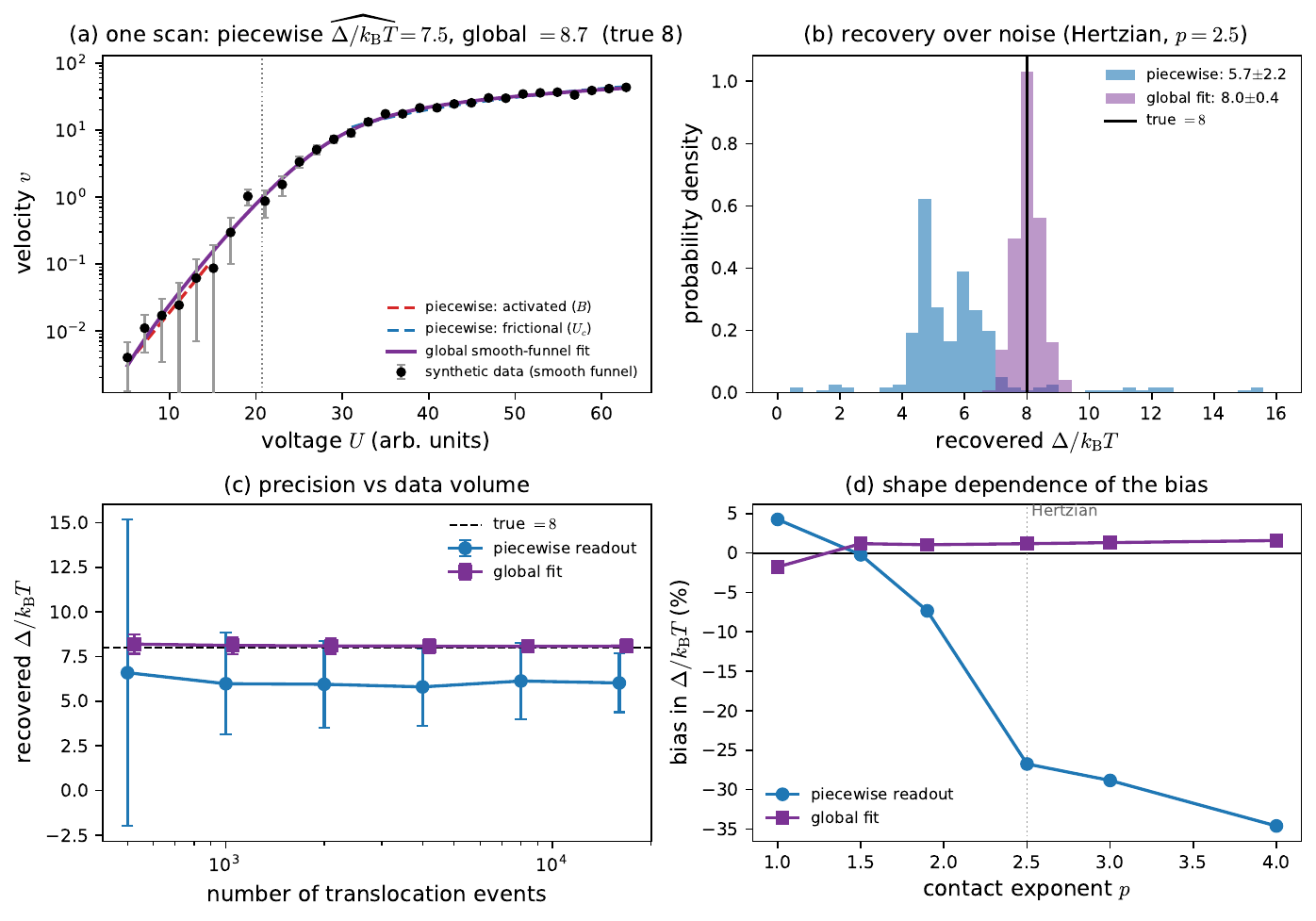}
  \caption{Translocation elastometry on synthetic data. (a) One
  velocity--voltage scan generated from a smooth (Hertzian) contact energy: the piecewise two-scaling
  fits (activated slope $B$, frictional intercept $\Uc$) underestimate $\Delta/\kT$, while a global
  fit of the whole curve to the smooth-funnel velocity---with the force-to-voltage conversion $c$
  free---recovers it. (b) Distribution of the recovered $\Delta/\kT$ over noise on the smooth data:
  the two-scaling readout is biased low, the global fit centred on the truth. (c) The two-scaling
  bias does not shrink with the number of events (it is structural), whereas the global fit
  converges. (d) Shape dependence: the two-scaling bias swings from slightly positive for near-linear
  ramps to strongly negative for smooth ones, while the global fit is unbiased across the contact
  exponent $p$.}
  \label{fig:recovery}
\end{figure*}

We test the theory with direct Langevin simulations of the model, using a vectorised
Euler--Maruyama integrator with the It\^o drift term appropriate to position-dependent diffusion
(Sec.~\ref{sec:numerics}). The mean velocity is measured from the steady-state source--sink current,
$v=J/\rho$, in exact correspondence with Eq.~\eqref{eq:vdef}. The simulated $v(\psi)$ agrees with the
closed form across the activated and frictional regimes to within a few percent---sub-percent in the
frictional regime, rising to a few percent at the most weakly driven, event-starved points
(Fig.~\ref{fig:sim}a, residual panel)---and the simulated translocation-time distribution matches the
analytic mean first-passage time~\cite{redner2001} to within a percent (Fig.~\ref{fig:sim}b).

As an in-silico stand-in for an experiment, we generate synthetic velocity--voltage data from the
exact theory with realistic, regime-dependent noise (events distributed in proportion to the local
rate, so the slow activated regime is sparsely sampled), and apply the elastometry extraction of
Eq.~\eqref{eq:elastometry}. When the data are generated from the model's own piecewise-linear
landscape, the injected deformation energy $\Delta/\kT=8$ is recovered with only a small positive
bias (a few percent) that decreases as the number of translocation events grows, the variance
falling as expected (Fig.~\ref{fig:recovery}).

The more demanding test is model misspecification: we generate the data from a \emph{smooth}
(Hertzian) contact energy---rising as the $5/2$ power of the indentation to the same plateau---but
fit them with the piecewise-linear scaling laws. Here the two-scaling readout is biased low by a
fixed amount, $\Delta/\kT\approx6.0$ for a true value of $8$ (a $\approx-26\%$ systematic), and this
bias does \emph{not} shrink with data volume: as events accumulate the estimate concentrates on the
biased value rather than the truth. The bias is shape dependent and non-monotonic in the contact
exponent: for nearly linear profiles it is small and slightly \emph{positive} (a few percent up to
$p\approx1.5$), changes sign near $p\approx1.9$, and then grows increasingly negative for smoother
profiles, reaching $\approx-26\%$ for the Hertzian case ($p=5/2$) and $\approx-40\%$ for a quartic one. Its origin is
structural and instructive. The readout Eq.~\eqref{eq:elastometry} is exact for the piecewise-linear
funnel precisely because that landscape produces a \emph{constant} restoring force in the entry
region, so the high-drive velocity is affine, $\tilde v\simeq\phi-\phi_c$, and its linear
extrapolation returns the true crossover force. For any smooth profile the entry force instead builds
up gradually, the high-drive gap decays as $\phi-\tilde v\propto\phi^{1-p}$ (exponent $p$ set by the
contact law), no sharp force threshold exists, and the extrapolated intercept $\Uc$ underestimates
$\Delta/(c\Lf)$---which is exactly where the bias enters.

This diagnosis points to a remedy that preserves the self-calibrating spirit of the method: rather
than extrapolate a straight line, fit the full scan to the exact velocity of a smooth funnel,
$\tilde v(\phi;\delta,\lambda,p)$, keeping the force-to-voltage conversion $c$ as a free parameter so
that no calibration is assumed. On the same synthetic Hertzian data this global-fit refinement
recovers $\Delta/\kT$ to within about $1.5\%$ at $N\simeq10^{3}$ events and to about $1\%$ or better for
larger samples ($N\gtrsim4\times10^{3}$), while also constraining the contact exponent $p$---though
$p$ is recovered with substantially larger relative uncertainty than $\Delta/\kT$ itself, which is
well determined because it depends only weakly on the assumed contact law. The two-scaling readout thus remains a fast, calibration-free estimate that
is correct to a good order of magnitude, while the global fit removes the misspecification bias when
the velocity curve is sufficiently sampled. The deformation energy is therefore an operationally
meaningful quantity, provided the shape sensitivity of the simple readout is recognised and, where
needed, corrected.

\section{Discussion}\label{sec:discussion}
The funnel model reduces deformable-particle translocation to a one-dimensional, exactly solvable
stochastic problem whose two dynamical regimes carry complementary information: the frictional regime
fixes the crossover force $\Psic=\Delta/\Lf$, the activated regime the funnel length $\Lf$, and their
combination the deformation energy $\Delta$ alone. Together with the geometric and contact-elasticity
maps, this turns a current--voltage measurement into a quantitative mechanical probe of single soft
particles---translocation elastometry---without immobilisation or labelling.

The model rests on assumptions that also delimit its scope: a single reaction coordinate (no internal
relaxation), a $z$-independent diffusivity inside the pore, and an elastic energy approximated as
piecewise linear. The misspecification test quantifies the cost of the last of these: the simple
two-scaling readout returns $\Delta/\kT$ correct to a good order of magnitude but with a systematic,
shape-dependent bias for smooth contact energies, which the global-fit refinement removes at the
price of a full-curve fit. Internal dynamics and hydrodynamic or electro-osmotic corrections to the
effective force are natural extensions. On the experimental side, the analysis suggests a concrete design rule:
sample the activated regime around $\psi\approx\Psic/2$, where the semilog slope is an unbiased
estimate of $\Lf/L$, and acquire enough low-drive events to control the variance of the readout.

Several directions extend the framework beyond the mean velocity used here. First, the passage is a
full stochastic trajectory, and the fluctuations of the current---not only its mean---carry
information about the local force landscape; inferring the effective force field directly from
recorded trajectories, as stochastic force-inference methods now allow~\cite{frishman2020}, could
turn a single translocation into a richer mechanical fingerprint. Second, the same funnel picture
applies well beyond solid-state nanopores: soft cells squeezing through microfluidic constrictions
are deformed and slowed in an analogous way~\cite{gambhire2017}, so the two-regime readout
may offer a label-free mechanical marker for red blood cells~\cite{atwell2022} and other soft biological objects.
Finally, forced translocation is central to cell migration, where a nucleus is driven through narrow
constrictions~\cite{amiri2024}; there the drive is typically strong enough that passage is
frictional rather than activated, a regime our closed form also covers, suggesting the same geometry
$\to L_f$ mapping could help interpret migration through pores. We hope this framework provides a
quantitative foundation for using pores---nanoscale and larger---as single-particle mechanical
sensors.

\begin{acknowledgments}
The project leading to this publication has
received funding from France 2030, the French Government
program managed by the French National Research Agency
(ANR-16-CONV-0001) and from Excellence Initiative of
Aix-Marseille University—A*MIDEX. P. R. thanks ICTP-
SAIFR (FAPESP Grant No. 2021/14335-0) where part of
this work was done. Co-funded by the European Union
(ERC-SuperStoc-101117322).
\end{acknowledgments}

\appendix

\section{Numerical methods}\label{sec:numerics}
All figures rest on the exact closed form Eq.~\eqref{eq:vfinal} except Figs.~\ref{fig:sim}
and~\ref{fig:recovery}, which use direct simulation and synthetic-data recovery. In reduced
coordinates $\zeta=z/L\in[-1,1]$ the reduced velocity is $\tilde v=1/B$ with $B=\int_{-1}^{1}
I(\zeta)\,\dd\zeta$ and integrating factor $I=\exp[v_{\mathrm{red}}(\zeta)-(1+\zeta)\phi]$,
$v_{\mathrm{red}}=V/\kT$; the three region integrals have removable singularities at
$\phi=0,\pm\phi_c$, taken by their analytic limits. We confirm the closed form against a
first-principles high-precision quadrature of $I(\zeta)$ over a dense grid in $(\phi,\delta,\lambda)$,
including at the removable singularities, with maximum relative error below $10^{-7}$.

\paragraph{Langevin simulation (Fig.~\ref{fig:sim}).} We integrate the reduced overdamped dynamics
$\dd\zeta=[D_r(\phi-v_{\mathrm{red}}'(\zeta))+D_r']\,\dd\tau+\sqrt{2D_r}\,\dd W$ in reduced time
$\tau=tD/L^2$ by explicit Euler--Maruyama, retaining the It\^o drift $D_r'$ so that a
position-dependent diffusivity does not bias the dynamics (we use constant $D_r=1$, the assumption
under which Eq.~\eqref{eq:vfinal} holds). The reduced force is piecewise constant---$\phi-\phi_c$ in
the entry funnel, $\phi$ inside, $\phi+\phi_c$ in the exit funnel. The mean velocity is the
source--sink steady state $v=J/\rho$: walkers exiting at $\zeta=+1$ reinject into a short reservoir
slab at the entrance and reflect at its far wall, so that $\rho$ is read at $\zeta=-1$ free of
reinjection pile-up and $J$ is the per-walker exit rate; a Brownian-bridge correction accounts for
mid-step crossings. Figure~\ref{fig:sim} uses $\delta=8$, $\lambda=0.5$, $3000$ walkers, and a
two-step Richardson extrapolation in the time step ($\dd\tau=5\times10^{-4}$ and $2.5\times10^{-4}$),
with error bars from eight walker sub-batches. Translocation times are first-passage times
$\zeta=-1\to+1$, pooled equally per walker to avoid length bias, and compared with the analytic mean
first-passage time.

\paragraph{Synthetic elastometry (Fig.~\ref{fig:recovery}).} Synthetic scans are drawn from the
theory---or, for the misspecification test, from a smooth Hertzian funnel whose elastic energy rises
as the $5/2$ power of the indentation to the same plateau---with multiplicative log-normal noise of
relative width $1/\sqrt{N_i}$, the events $N_i$ at each voltage distributed in proportion to the local
rate (so the activated regime is sparsely sampled). The two-scaling readout fits the activated
log-slope in a window about $\Uc/2$ and the frictional line above $1.5\,\Uc$, with $\Uc$ located by a
blind first pass. The global-fit refinement fits the full scan to the smooth-funnel reduced velocity
$\tilde v(\phi;\delta,\lambda,p)$, obtained by quadrature of $I(\zeta)$ for the smooth profile, with
$(\delta,\lambda,p,c)$ all free. Recovery uses $\delta=8$, $\lambda=0.4$, $c=0.7$, and event counts
$N$ from $5\times10^{2}$ to $1.6\times10^{4}$ over several hundred noise realisations per point. Every
figure is regenerated from a single script.

\bibliographystyle{apsrev4-2}
\bibliography{refs}

\end{document}